# Collimation of a Circulating Beam in the U_70 Synchrotron by Use of Reflections in Axially – Oriented Crystals.

A. G. Afonin, V. T. Baranov, M. K. Bulgakov, I. S. Voinov, V. N. Gorlov,
I. V. Ivanova, D. M. Krylov, A. N. Lunn'kov, V. A. Maisheev, S. F. Reshetnikov,
D. A. Savin, E. A. Syshchikov, V. I. Terekhov, Yu. A. Chesnokov,
P. N. Chirkov, and I. A. Yazynin

*Institute for High Energy Physics, Protvino, Moscow region, 142281 Russia*

Abstract:
The possibilities of the extraction and collimation of a circulating beam by a new method due to the reflection of particles in crystals with axial orientation were experimentally investigated in the Fall-2010 run at the U_70 synchrotron. Such crystals have positive features, because the axial potential is five times larger than the planar potential. It has been shown that the collimation efficiency can reach 90% due to axial effects in the crystal. Losses of the circulating beam on a collimator have been reduced by several times; this makes it possible to suppress the muon jet near the steel collimator of the circulating beam.

Scientists working at the Institute for High Energy Physics, jointly with colleagues from several Russian and foreign research centers, recently discovered a new physical phenomenon: the reflection of a high_energy proton beam from bent atomic planes of a silicon crystal [1–4]. Wolume reflection is caused by the interaction of an incident proton with the potential of the bent atomic lattice and occurs in a small length in the region tangential to the bent atomic plane, leading to the deflection of the particle towards the side opposite to the bend. The probability of the reflection is large and approaches unity at energies of about 100 GeV. Reflection occurs in a wide angular range and is more efficient than usual channeling [5]. For this reason, there are real prospects for using reflection in the extraction and collimation of beams at large accelerators.

The deflection angle of particles reflected from crystallographic planes is limited by 1.5 $\theta_c$ (see, e.g., [2,4]). Here, $\theta_c = (2U_c/pv)^{1/2}$ is the critical channeling angle, where $U_c \sim 20$ eV is the (111) planar potential in silicon and $p$ and $v$ are the momentum and velocity of the incident particle. Accelerator physics problems require an increase in the reflection angle by several times. To this end, the following two methods were proposed: first, reflection on a chain of crystals [6] and, second, reflection in one crystal oriented near the axis by the total potential of several skew planes [7, 8].

The principle of an increase in the reflection angle on the chain of crystals is illustrated in Fig. 1. Using several ($N \leq 10$) successively located oriented crystals, the particle deflection angle can be increased by a factor of 5–10. A further increase in the number of crystals leads to a decrease in the efficiency of the process in view of an increase in the number of nuclear interaction events.

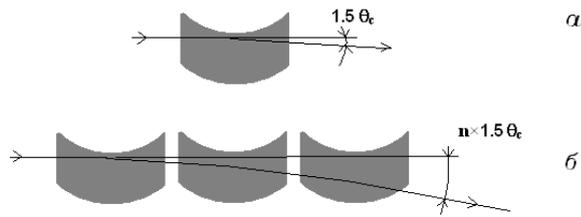

**Fig. 1.** Increase in the reflection angle on a chain of crystals.

Figure 2 shows an increase in the particle deflection angle oriented near the crystallographic axis. The summarized effect of the reflection from several skew planes can increase the particle trajectory deflection by a factor of $A_F \sim 3\text{-}5$ ($A_F$ is the axial factor).

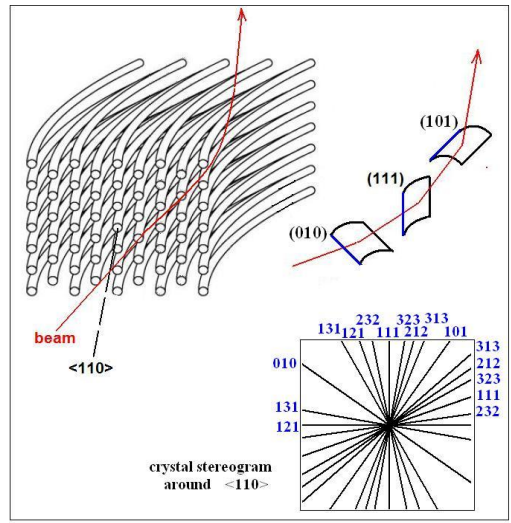

**Fig. 2.** Scheme of the deflection of particles near the crystallographic axis <110>: the effect from a main bent (111) plane is supplemented by reflection from skew (010) and (101) planes and other, weaker planes.

An increase in the reflection angle on a chain of ten bent crystals was used at the U_70 synchrotron to improve the collimation of the beam [9]. To the same end, a chain of eight oriented crystals was used at the Tevatron accelerator [10]. In this work, we demonstrate the most optimal method for increasing the particle deflection angle in a crystal due to reflection by simultaneously using two mentioned methods; a chain of six bent crystals is oriented near the <110> crystallographic axis. Thus, the proton reflection angle is increased by a factor of $(N \times A_F) \sim 20$ as compared to the process in a single bent crystal with the planar orientation.

To implement such a crystal device, we used a design in Fig. 3a. The crystals were bent due to the anisotropic properties of the crystal lattice. A necessary transverse bend of 1 mrad appears when each strip is bent in the longitudinal direction. Furthermore, since all strips are made from one silicon plate, the ideal mutual orientation of individual strips holds in both the horizontal and vertical planes. This property makes it possible to mount a multideflector in a biaxial goniometer and to ensure the deflection of particles in both the planar and axial cases (see Fig. 3b).

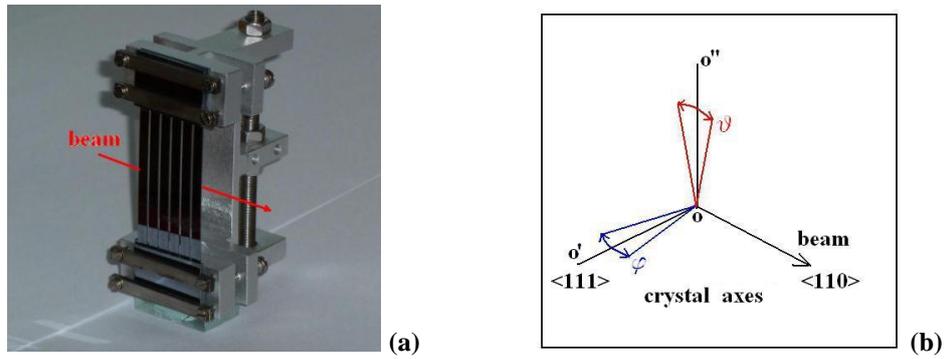

**Fig. 3.** Design of a silicon multideflector (consisting of six crystals) and the scheme of its rotation in a biaxial goniometer.

The biaxial goniometer with the multicrystal was placed in the vacuum chamber of the accelerator in the system of the localization of losses of the U_70 accelerator at a certain distance in front of the steel absorber to ensure the necessary throw of the beam (see Fig. 4). The efficiencies of the operation of the mounted crystals and the collimation system were determined by profilometers at the end of the absorber and the ionization chambers located downstream of the absorber.

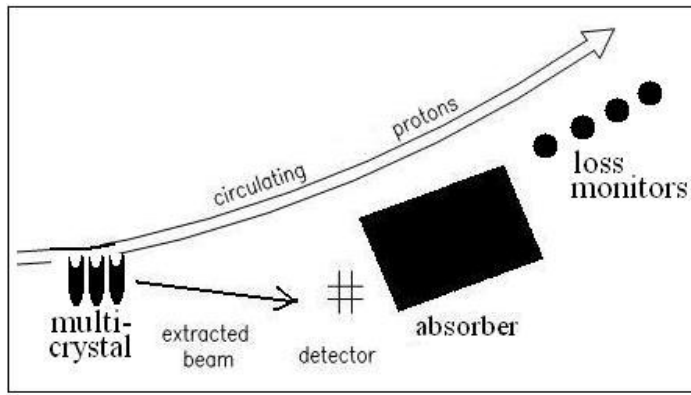

**Fig. 4.** Scheme of the experiment with a multicrystal on the circulating beam of the U_70 synchrotron.

In the experiment, the 50_GeV proton beam was guided to the multicrystal by means of the horizontal distortion of the closed orbit by a smoothly increasing bump. After the beam_guiding adjustment, the multicrystal was introduced to the regime of the reflection of particles on the (111) planar potential by means of the horizontal rotation. Then, the vertical angle of the goniometer was varied. Figure 5a shows the depth of the throw particles on the absorber (distance measured from the front edge) as a function of the vertical rotation angle of the goniometer. The throw enhancement effect, which is due to the axial potential of the crystal and coincides with the predicted value, is clearly seen. The passage of particles in the reflection regime through six bent crystals was simulated by the Monte Carlo method using the SCRAPER software (see Fig. 5b) [11].

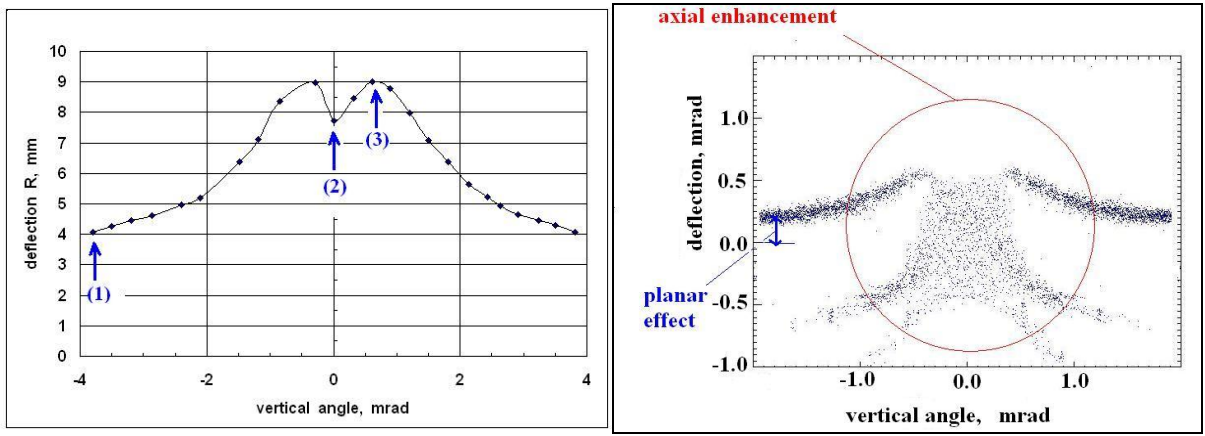

(a) (b)

**Fig. 5.** (a) Depth of the throw of 50 GeV protons on an absorber versus the vertical rotation angle. (b) Monte Carlo calculation of an increase in the angle of the reflection of particles in the multireflector due to axial effects.

Figure 6 shows a decrease in losses of particles at the accelerator downstream of the absorber, measured by four monitors under the variation of the horizontal and vertical rotation angles.

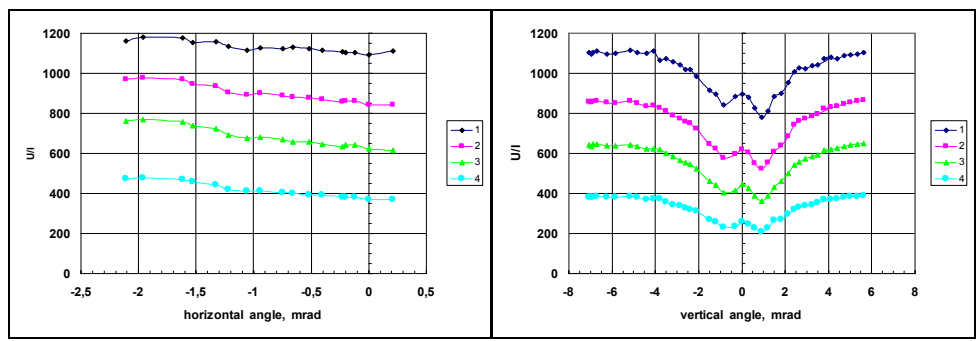

(a) (b)

**Fig. 6.** (a) Decrease in losses under the rotation of the horizontal angle (planar reflection). (b) Further decrease in losses due to the adjustment of the vertical angle of the goniometer (reflection near the axis).

Figure 7 shows the characteristic beam profiles on the end of the absorber collimated with and without the crystal. It is seen that the maximum throw of the beam and, correspondingly, the improvement of collimation occur with the application of the reflection of particles in the case of the axial orientation of the crystal deflector.

The efficiency of the beam throw onto the absorber reached 90%. The determination of the collimation efficiency and other experimental details were described in [12].

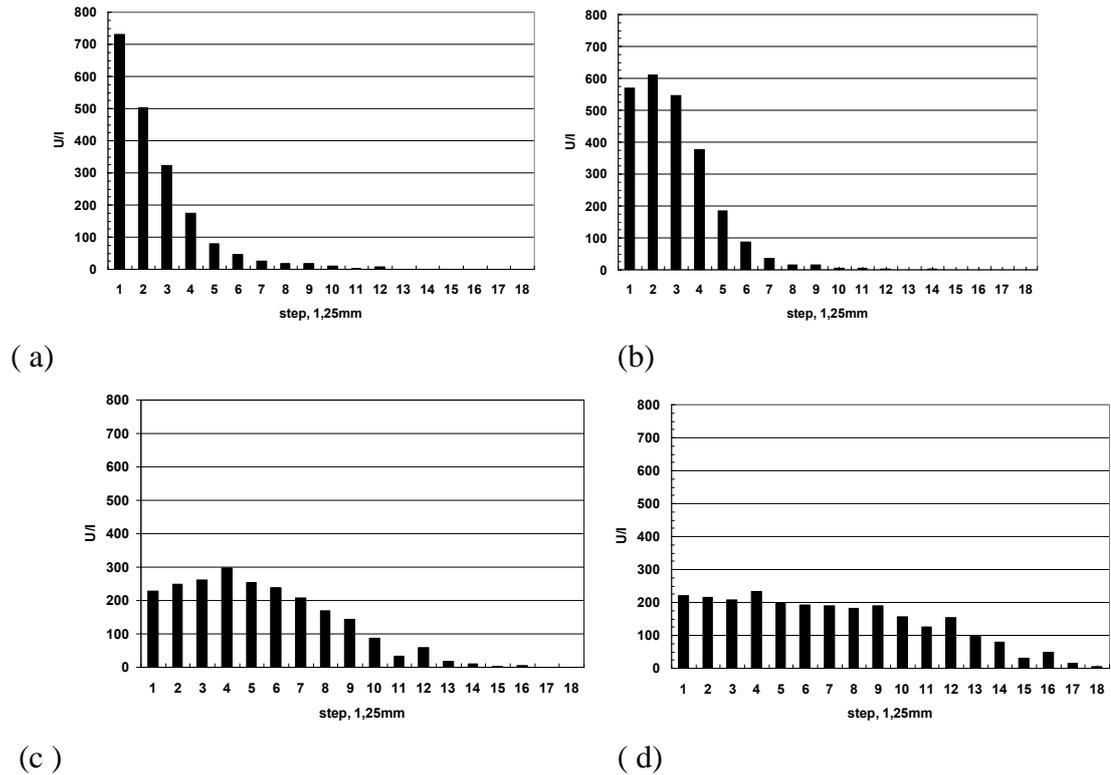

**Fig. 7.** Beam profiles on the end of the absorber in the case of (a) amorphous scattering, (b) reflection from the atomic planes (angular position (1) in Fig. 5a), and (c, d) reflection near the crystal axis (angular positions (2) and (3) in Fig. 5a, respectively).

To summarize, we have described the method for controlling the trajectories of particles based on the reflection of particles in the multicrystal enhanced by the axial effects. It has been shown that losses of the beam under the optimal adjustment of the multicrystal decrease by a factor of 2 or 3 as compared to the disordered case, in agreement with the calculation. The application of crystals with the axial orientation is also promising for the collimation of beams at the Large Hadron Collider and International Linear Collider. This work was supported by the Directorate of the Institute for High Energy Physics, by the Rosatom Nuclear Energy State Corporation (contract no. N.4e.45.03.09.1047), and by the Russian Foundation for Basic Research (project no. 08_02_01453_a).